\documentclass[aps,prl,twocolumn,showpacs]{revtex4}

\usepackage{graphicx}

\begin{document}

\title{Tunable  conductance of  magnetic nanowires with  structured
domain walls}

\author{V. K. Dugaev$^1$}
%\email{vdugaev@cfif.ist.utl.pt}
\author{J. Berakdar$^2$}
\author{J. Barna\'s$^{3,4}$}
\affiliation{$^1$Departamento de F\'isica and CFIF, Instituto Superior T\'ecnico,
Av. Rovisco Pais, 1049-001 Lisboa, Portugal\\
$^2$Max-Planck-Institut f\"ur Mikrostrukturphysik,
Weinberg 2, 06120 Halle, Germany\\
$^3$Department of Physics, Adam Mickiewicz University, Umultowska~85, 61-614~Pozna\'n, Poland\\
$^4$Institute of Molecular Physics, Polish Academy of Sciences,
Smoluchowskiego~17, 60-179~Pozna\'n, Poland}

\date{\today }

\begin{abstract}
We show that in a   magnetic nanowire
with  double
magnetic domain walls, quantum interferences result
in spin-split quasistationary  states
 localized mainly in between the domain walls.
Spin-flip-assisted transmission
through the domain structure increases strongly  when these
 size-quantized states are tuned on resonance with the Fermi energy, e.g. upon
 varying the distance between the  domain walls which results  in
 resonance-type peaks  of the  wire conductance. This novel phenomena is shown to be
 utilizable  to manipulate
 the spin density in the domain vicinity.
The domain walls parameters are readily controllable  and  the
predicted effect is hence exploitable in
 spintronic devices.
%
%We propose a new effect of resonant conductance of a magnetic
%nanowire with a double domain wall. The effect is related to the
%quantum interference of electron wave function in the spin quantum
%well, and results from the quantization of electron energy states
%in the spin quantum well created by the double domain wall. The
%spin-flip-assisted transmission through the domain wall is
%effective only when the level energy in the spin quantum well is
%close to the Fermi energy. This leads to a resonance-like
%dependence of the magnetic wire conductance on the distance
%between the domain walls. Since the distance between the walls can
%be easily controlled, the effect can be used in novel spintronic
%devices.
\end{abstract}
\maketitle
The discovery of the giant magnetoresistance \cite{gmr} and its
rapid and diverse  industrial utilizations sparked major efforts
in understanding and exploiting  spin-dependent transport
phenomena. In particular, new perspectives of even broader
importance are anticipated from combining semiconductor technology
with nanoscale fabrication techniques to produce magnetic
semiconducting materials and control precisely the spins of the
carriers. At the heart of this new field that is now termed
"semiconductor spintronics" \cite{prinz98} is the understanding of
the transport properties of a magnetic domain wall (DW), which is
a region of inhomogeneous magnetization in between two domains of
 homogeneous (different) magnetizations.  Thick (or adiabatic) DWs
 occurring  in bulk metallic ferromagnets  have an extension
much larger than the carriers Fermi wave length and are largely
irrelevant for the  resistance \cite{cabrera74}. In contrast, a
series of recent experiments on magnetic nanostructures,
 and particularly nanowires, revealed
that the magnetoresistance  in the presence of
DWs can be as large as  several
hundreds \cite{garcia99,ebels00} or even
thousands \cite{chopra02,ruster03} of percents.
 These observations have decisive consequences in so far as
 DWs are controllable efficiently  by  applying a magnetic
field \cite{yamaguchi04} and can also be steered  by  a
spin-polarized electric current \cite{katine00}, meaning that the
magnetoresistance of the structure containing DW is  controllable
via an electric field.

%For good metallic ferromagnets such as  Ni
%many facets of the underlying
% physics of this phenomenon are still unclear
%\cite{simanek01}.
% On the other hand,

  The interpretation of  the  huge magnetoresistance of DWs observed
  in magnetic semiconductor
(in the ballistic  quantum regime) relies on the relative
sharpness of DWs on the scale set by the wavelength of carriers
(electrons or holes)
\cite{kudrnovsky00,tagirov01,yavorsky02,dugaev03,zhuravlev03}. In
such a situation spin-dependent scattering of  carriers from DWs
is greatly enhanced. In this paper we predict the formation of
spin quantum wells and the occurrence  of a novel effect in
magnetic semiconductor nanowires with double DWs:  By pinning one
of the DWs at a constriction, one can control the location of the
second DW. The spin-dependent transmission and reflection of
carriers waves from the first and the second DW and the quantum
interference between these waves lead to the formation of
spin-split quasistationary states and hence the double DWs act in
effect as a penetrable "spin quantum well" (located in between the
two DWs). Lifetimes and  energetic positions of these states
depend on experimentally controllable parameters of  DWs  such as
the extent of the well, i.e. on the distance between DWs. As a
result, the conductance of a structure with  DWs possesses sharp
resonances when the Fermi energy matches the spin quantum-well
states. Hence, the resistance varies by several orders in
magnitude in response to minor changes in  DW positions or in
carrier density, e.g. achieved upon gating the structure. Strong
backscattering from DWs and interferences between  carrier waves
lead to spin-density accumulation that can be tuned externally by
modifying the DWs structure.

We consider a magnetic wire with a magnetization profile exhibiting
two DWs separated by the distance
$2L$. The magnetization vector field ${\bf M}(z)$ in both DWs varies within
the $x-z$ plane. The $z$ axis is along the wire. Thus,
$z$ is the easy axis, and the $x-z$ plane is the easy plane.
 We study the case where the thickness and the width of the wire are
 smaller than the carrier Fermi wavelength so that only one size quantized level
(a single one-dimensional subband) is populated.
This situation is   realizable  in magnetic semiconductor-based structures.
\begin{figure}
%\hspace*{-1cm}
\includegraphics[scale=0.35]{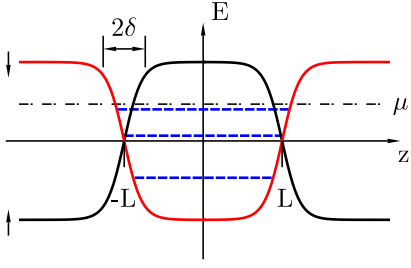}
\vspace*{-0.5cm}
\caption{(Online color) Schematics of the double
domain wall structure: indicated are the potential energy profiles
for spin-up and spin-down electrons. Dashed lines between domain
walls stand for quasi-localized energy levels in the spin-down
quantum well.}
\end{figure}
The Hamiltonian describing independent carriers  along the wire in  presence of the
magnetization field ${\bf M}(z)$  is
\begin{equation}
\label{1}
H=-\frac{\hbar ^2}{2m}\, \frac{d^2}{dz^2}-JM_z(z)\,
\sigma _z-J M_x(z)\, \sigma _x\; ,
\end{equation}
where $J$ is the exchange coupling constant, $M_{x(z)}$ is the
$x\, (z)$ component  of  the inhomogeneous magnetization field, and $m$ is the carriers effective mass.
Figure~1 shows a  schematic drawing  of the spin up
and spin down band-edge profiles
that are utilized for a full quantal treatment of the spin-dependent scattering of charge carriers.

We are interested in the case where  the width $2\delta $ of each DW  is  smaller than
the carriers Fermi wavelength, $k_F\delta \ll 1$ and particularly
 when $k_FL\geq 1$ (otherwise the carriers are not
influenced by the  detailed topology of the DWs).
For moderate carriers density
the chemical potential $\mu $ is in one of the magnetically
split subbands (cf. Fig.~1). This case
corresponds to a full spin polarization of the electron gas, a situation
 realizable in magnetic semiconductor nanowires.
Carriers wave functions   are expressible as
\begin{eqnarray}
\label{2-4}
\psi _{k}(z) =\left( e^{ikz}+r\, e^{-ikz}\right) \left|
\uparrow \right> +r_f\, e^{\kappa z}\left| \downarrow \right> ,
\hskip0.5cm z<-L,\hskip0.2cm \\
\psi _{k}(z)=\left( A\, e^{\kappa z}+B\, e^{-\kappa
z}\right) \left| \uparrow \right> +\left( C\, e^{ikz}+D\,
e^{-ikz}\right) \left| \downarrow \right> ,
\nonumber \\
|z|<L,\hskip0.2cm \\
\psi _{k}(z) =t\, e^{ikz}\left| \uparrow \right> +t_f\,
e^{-\kappa z}\left| \downarrow \right> , \hskip0.5cm z>L,\hskip1cm
\end{eqnarray}
where
 $\left| \uparrow \right>$ ($\left| \downarrow \right>$) is
the spin-up  (spin-down) component of the carrier states,
$k=[2m(\varepsilon +JM)]^{1/2}/\hbar $, and $\kappa
=[2m(JM-\varepsilon )]^{1/2}/\hbar $. The electron energy
$\varepsilon $ is measured from the midpoint in between  spin-up
and spin-down band edges. The  non-spin-flip (spin-flip)
transmission and reflection coefficients $t$ and $r$ ($t_f$ and
$r_f$) as well as the constants $A$, $B$, $C$ and $D$ have to be
deduced from the solutions of Eqs.~(\ref{1}) and from wave
function continuity requirements.
 Physically, Eqs.~(2)-(4)
describe  spin-up carriers incoming from the left, being
transmitted and reflected from the double DW structure into
waves with the same or opposite spin polarizations with a
subsequent decay of the spin-down part of the wave function.

To determine the unknown coefficients in Eqs.~(2)-(4)
we utilize the wave function continuity  at $z=\pm L$, i.e.
\begin{equation}
\label{5}
\frac{\hbar }{2m}\left( \left. \frac{d\psi
_{k}}{dz}\right| _{z=-L+\delta } -\left. \frac{d\psi
_{k}}{dz}\right| _{z=-L-\delta }\right) +\lambda\,\sigma _x \,
\psi _{k}(-L)=0,
\end{equation}
where
\begin{equation}
\label{6}
\lambda\simeq \frac{J}{\hbar } \int _{-L-\delta
}^{-L+\delta } dz\; M_x(z)\simeq \frac{2JM\delta }{\hbar }\; .
\end{equation}
Similar equation holds for $z=L$.
The boundary conditions at $z=\pm L$ (eight equations for the
spinor components) define all the coefficients in Eqs.~(2)-(4), e.g.,
%One can reduce the number of equations by excluding the
%coefficients $A,\, B,\, C,\, D$, which determine the wave function
%in the region between the two DWs. Then, one obtains the following
%set of equations
\begin{eqnarray}
\label{7}
\left[ \kappa -ik\, \tanh \, (2\kappa L)\right]
\tilde{r} -\Delta \, \tanh \, (2\kappa L)\; \tilde{r}_f\hskip0.5cm
\nonumber \\
-\frac{\kappa \, \tilde{t}}{\cosh \, (2\kappa L)} =-\left[ \kappa
+ik\, \tanh \, (2\kappa L)\right] e^{-ikL},
\end{eqnarray}
\begin{eqnarray}
\label{8}
-\Delta \, \sin \, (2kL)\; \tilde{r} +\left[ \kappa \,
\sin \, (2kL)+k\, \cos \, (2kL)\right] \tilde{r}_f
\nonumber \\
-k\; \tilde{t}_f
=\Delta \, \sin \, (2kL)\; e^{-ikL},
\end{eqnarray}
\begin{eqnarray}
\label{9}
-\left[ \kappa \, \tanh \, (2\kappa L)-ik\right]
\tilde{r} +\Delta \, \tilde{r}_f +\frac{ik\; \tilde{t}}{\cosh \,
(2\kappa L)}\,
\nonumber \\
+\frac{\Delta \; \tilde{t}_f}{\cosh \, (2\kappa L)}\,
=\left[ \kappa \, \tanh \, (2\kappa L)+ik\right] e^{-ikL},
\end{eqnarray}
\begin{eqnarray}
\label{10}
\Delta \, \cos \, (2kL)\; \tilde{r} -\left[ \kappa \,
\cos \, (2kL)-k\, \sin \, (2kL)\right] \tilde{r}_f
\nonumber \\
+\Delta \; \tilde{t}
-\kappa \; \tilde{t}_f
=-\Delta \, \cos \, (2kL)\; e^{-ikL}.
\end{eqnarray}
Here we introduced the following notation: $\Delta =2m\lambda
/\hbar $, $\tilde{r}=r\, e^{ikL}$, $\tilde{r}_f=r_f\, e^{-\kappa
L}$, $\tilde{t}=t\, e^{ikL}$, and $\tilde{t}_f=t_f\, e^{-\kappa
L}$. In this way  we derived (rather cumbersome) analytical
expressions   for the reflection and
transmission coefficients.
\begin{figure}
\vspace*{-1cm}
\hspace*{-1cm}
\includegraphics[scale=0.4]{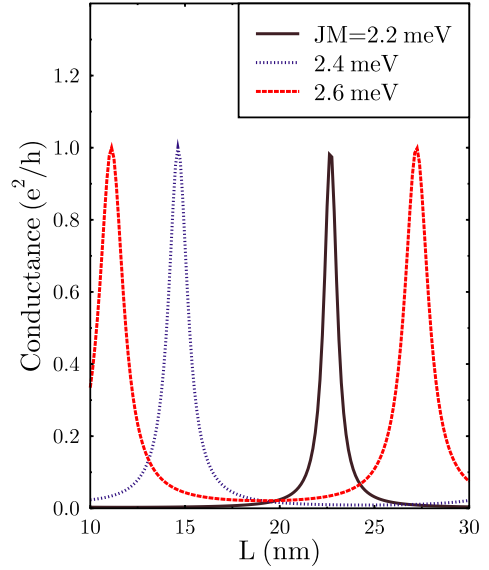}
\vspace*{-0.6cm}
\caption{(Online color) Conductance vs. distance between two domain walls for
different values of the magnetization $M$.
For the numerical calculations  we assumed
$m=0.6\, m_0$ ($m_0$ is the free electron mass), $\delta =2$~nm, and $\mu
=-2$~meV. }
\end{figure}
\begin{figure}
\vspace*{-1cm}
\hspace*{-1cm}
\includegraphics[scale=0.4]{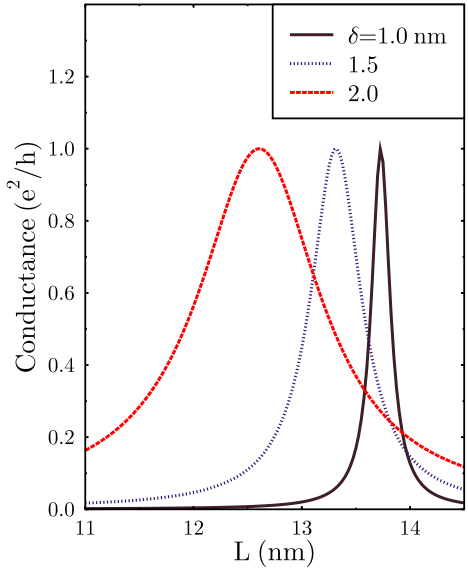}
\vspace*{-0.6cm}
\caption{(Online color) Conductance vs. distance between two domain walls for
different values of the domain wall width  $\delta $. Other parameters are:
$JM=2.5$~meV, $m=0.6\, m_0$  and
$\mu =-2$~meV.}
\end{figure}

For a physical insight into the results we inspect at first
 the limiting situation  $\Delta ={4m JM\delta }/{\hbar^2 }=0$ in which case
 no spin-flip transitions  occur  at
the DW. Correspondingly, only spin-up electrons tunnel
through the  barrier.
The standard formula for  barrier tunnelling
%
%\begin{equation}
%\label{11}
%
$t={2ik\kappa \, e^{-2ikL}} [2ik\kappa \, \cosh \,
(2\kappa L)+(k^2-\kappa ^2)\, \sinh \, (2\kappa L)]^{-1}\; $ is
retrieved using Eqs.~(7),(8).
%\end{equation}
%The corresponding contribution to the conductance is very small,
%and for $\kappa L\gg 1$ it can be completely neglected.
%
The spin-down electrons are localized within the spin quantum
well. From Eqs.~(9) and (10) we  find the
symmetric solution (with $r_f=t_f$) that corresponds to  localized
states with the wave vector $k$ and obeys  the relation $\tan (kL)=\kappa
/k$. The antisymmetric solution for such  $k$ (with $r_f=-t_f$)
satisfies the equation
$\tan (kL)=-\kappa /k$. When  the distance $L$ between
the DWs is varied the energetic positions of the
 size-quantized levels within the well
are shifted. For certain values of $L$ the energy  of the localized states
within the well (dashed lines in Fig.~1) coincide with the Fermi
level. Thus, if the spin-mixing amplitude is finite (i.e. $\Delta \ne 0$)
we expect spin-up carriers   to  transverse  resonantly the DWs.
It is important to note here  that when $\Delta \ne 0$,
the aforementioned localized spin-down states (dashed lines in Fig.~1)
 turn quasi stationary with  a finite decay width
$\Gamma =\hbar /\tau \sim |t_f|^2$ where  $\tau $ is the life time
of these quasilocalized states.  $|t_f|$ is controllable, e.g.
by changing the parameters of the DWs.

In the regime of a linear response to an applied
bias voltage  the conductance of the wire is determined
by the transmission coefficient $t$ as
\begin{equation}
\label{11}
G=\frac{e^2}{2\pi \hbar }\; \left| t(\varepsilon =\mu) \right| ^2.
\end{equation}
 Using this relation and the  solutions derived from Eqs.~(7)-(10) we
calculated the variation of the conductance $G$ with the DWs distance
(2$L$) for several values of the magnetization $M$.
The  conductance shown in Fig.~2 exhibits narrow
resonance peaks corresponding to those values of $L$ at which a
quasi-discrete, size-quantized level coincides with the Fermi
energy. At the conductance peak position the effective barrier created by
the DWs is basically transparent.
The width of the resonance peaks
is related to $\tau$, the  life time of the quasi-stationary, spin-well
states and  is determined by the  spin-mixing mechanism  mentioned above
(and specifically by $|t_f|$). As demonstrated by
Fig.~3, the strength of spin mixing and hence the width of the resonance peaks
can be controlled, e.g., by varying the width $\delta $ of the DWs.
Decreasing the
spin-mixing parameter $\Delta = {4m JM\delta }/{\hbar^2 }$,
the life time of the localized  spin quantum-well states increases
and the conductance resonance peaks become correspondingly narrower.
%The transmission is 100\% for electron energies
%coinciding with the energy of discrete levels. In this case of a full
%resonance, the effective potential is reflectionless.
%(\emph{why the value of $G$ is not changed! when $\Delta \to 0$} !!)
The energetic positions of the quasi discrete levels  depend also on
$\Delta $ (and hence on $\delta$). This
 results in a slight shift of the resonance positions  when changing $\delta$, as
 shown in Fig.~3.
Experimentally  the Fermi level position can be shifted by
electrically gating  the whole structure. In this case,
the resonance conductance peaks  occur
as a function of the gate voltage for a fixed distance
between the DWs.

\begin{figure}
\vspace*{-1cm}
\hspace*{-1cm}
\includegraphics[scale=0.4]{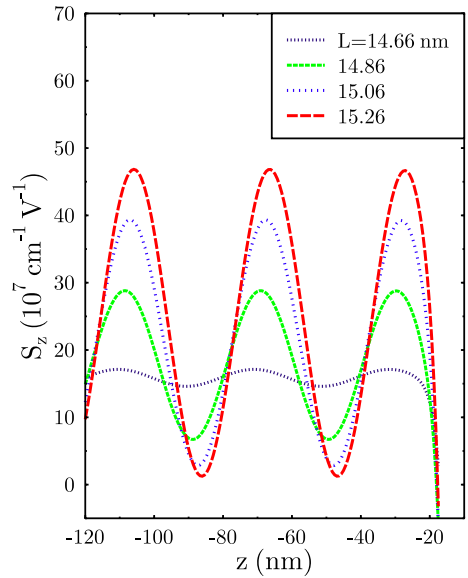}
\vspace*{-0.6cm}
\caption{(Online color) Spin density profile for $z<-L$ at different values of
spin quantum well $L$. Here $S_z$ is the density of electron spins
per unit wire length and unit voltage bias.
The parameters are $JM=2.4$~meV, $\mu =-2$~meV, and $\delta =2$~nm.}
\end{figure}

It is important to remark that within our model  the resonant transmission  does
not vanish in the limit of large $L$, for
we do not incorporate
 decoherence effects that destroy the interference of
transmitted and reflected waves in the region in between the magnetic
DWs. Thus, our  present considerations are valid for $L\ll L_e$,
where $L_e$ is the decoherence length. On the other hand we expect the double DWs resistance
in this limit to be the sum of the resistances of the individual DWs which we calculated
previously \cite{dugaev03}.
Furthermore, we  remark that while our calculations are done for $T=0$,
the effect of the temperature $T$ is negligibly small as long as $T\ll
E_0$, where $E_0$ is the effective barrier for spin-up electrons,
$E_0=JM-\mu $. For higher temperatures, the activated spin-up
electrons contribute to the conductance, and the resonant
character of the conductance is smeared out.

A further notable feature of the scattering of spin-split carriers
from DWs is the interference-induced buildup of spin-density in
the DWs vicinity, as demonstrated in Fig.~4. The period of the
spin-density oscillation are determined by the Fermi wave vector
(close to the DWs  there is a fast-decaying contribution
(proportional to $e^{\kappa z}$) which is superimposed on the
simple oscillations  shown in Fig.~4). The degree of the
spin-density accumulation can be controlled, to a certain extent,
by changing experimentally the parameters of the DWs. For
instance, for the well size $L\simeq 14.66$ nm quasistationary
well states are formed and are energetically
 close to the Fermi energy. As a result  DWs are almost non-reflecting
and the  spin-density buildup diminishes (cf. Fig.~4). Changing
$L$ (i.e., off resonance)  DWs backscatter strongly and large
spin-density  is accumulated.
%(the phase shift of the oscillation
%observed with increasing $L$ is due to the phase shift of the
%scattered carrier waves  when going through the resonance).
 By gating the structure we can tune
the Fermi energy and manipulate the spin density in a manner similar to Fig.~4.
For an experimental verification we note that
 spin-density modulations can be imaged   with a sub nanometer resolution
using spin-polarized scanning tunneling microscopy \cite{ute04}.
Hence, the  predictions of Fig.~4 are  accessible experimentally.

It is useful to compare the present findings with  resonant
tunnelling for  manipulating the spin orientation in magnetic
layered structures \cite{gruber01} in which case the discrete
levels are controlled {\it via} an external magnetic field that
modifies the spin splitting of quantum-well energy levels. In
contrast our proposition relies on a interference-induced creation
of a {\it spin quantum well} by the presence of double DW
structure and as demonstrated above offers a range of external
parameters with which the conductance can be tuned. For an
experimental realization nanowires of low-density magnetic
semiconductors \cite{matsukara} ($5\times 10^{18}$~cm$^{-3}$)
 are favorable. Such structures are reported in \cite{ruster03}, however
 the DWs separation was  too large (500~nm) for a
noticeable  effects. In principle however, the experiment should
be feasible with the parameters employed above: E.g., the wire
Fermi momentum is $k_F = \pi \rho_{1D}$, where $\rho_{1D}$ is the
linear hole density related to the bulk density by $\rho_{1D} =
\rho_{3D}\ S$ and $S$ is the  wire cross section. For $\rho_{3D}
= 10^{20}$~cm$^{-3}$ and $S = 1$~nm$^2$, we obtain
$k_F\simeq 3\times 10^6$~cm$^{-1}$ corresponding to a wavelength
$\lambda_F\simeq 20$~nm and an
energy $E_F\simeq 5$~meV. These numbers are in the range of those
used in our calculations.

Our considerations assumed electrons as carriers. In III-V
magnetic semiconductors the carriers  are
 holes with an energy spectrum  more complicated than that
described by Eq.~(1).
% The underlying physics remains the same, as
It is clear however that in this case carriers scattering and
inferences lead to the formation of a spin quantum well with
localized levels and hence the physical phenomena discussed above
are expected to emerge  as well for the case of  holes carriers.
For a strong magnetization and hence large splitting of the
valence subbands, $JM \gg |\mu |$ ($\mu $ is the chemical
potential measured from the valence band edge) one can employ a
model with parabolic bands as for electrons.

Summarizing, the conductance of a magnetic nanowire with a double
DWs is shown to possess  a strong dependence
 on the DWs separation. The
extreme sensitivity of the conductance on the
 inter-walls  distance can be used to identify the relative position of the
DWs. It can also be utilized to transform a magnetic field effect
 on the DWs into a change of the current flowing through
the nanowire. A wall displacement of the order of 10\% induces a
resistance change of hundreds of percents.
%We also predict a spin-density buildup whose experimental verification  will
%shed new light on the physics of transport through DWs.

This work is supported by FCT Grant POCI/FIS/ 58746/2004
(Portugal) and by Polish State Committee for Scientific Research
under Grants PBZ/KBN/ 044/P03/2001 and 2~P03B~053~25.

\end{document}